# External Electromagnetic Wave Excitation of a Pre-Synaptic Neuron Based on LIF model

Emad Arasteh Emamzadeh-Hashemi
School of Electrical and Computer Engineering
University of Tehran
Tehran, Iran
emad.arasteh@ut.ac.ir

Ailar Mahdizadeh
School of Electrical and Computer Engineering
University of Tehran
Tehran, Iran
ailar.mahdizadeh@ut.ac.ir

*Abstract*—Interaction of electromagnetic (EM) waves with human tissue has been a longstanding research topic for electrical and biomedical engineers. However, few numbers of publications discuss the impacts of external EM-waves on neural stimulation and communication through the nervous system. In fact, complex biological neural channels are a main barrier for intact and comprehensive analyses in this area. One of the ever-present challenges in neural communication responses is dependency of vesicle release probability on the input spiking pattern. In this regard, this study sheds light on consequences of changing the frequency of external EM-wave excitation on the post-synaptic neuron's spiking rate. It is assumed that the penetration depth of the wave in brain does not cover the post-synaptic neuron. Consequently, we model neurotransmission of a bipartite chemical synapse. In addition, the way that external stimulation affects neurotransmission is examined. Unlike multiple frequency component EM-waves, the monochromatic incident wave does not face frequency shift and distortion in dispersive media. In this manner, a single frequency signal is added as external current in the modified leaky integrated-and-fire (LIF) model. The results demonstrate existence of a node equilibrium point in the first order dynamical system of LIF model. A fold bifurcation (for presupposed LIF model values) occurs when the external excitation frequency is near 200 Hz. The outcomes provided in this paper enable us to select proper frequency excitation for neural signaling. Correspondingly, the cut-off frequency reliance on elements' values in LIF circuit is found. In addition, the spontaneous and stimulated vesicle release effects on post-synaptic neuron's spiking rate are computed. It will be shown that the neurotransmission probability and output spiking rate can be optimized based on proper selection of excitation frequency.

I. INTRODUCTION

Neural activities in nanoscopic scale dimensions of brain neurons can be modeled as a communication signaling procedure. Hence, spiking rates of the post-synaptic neurons can be analyzed like a noisy received signal. As a result, neural networks can be designed to simulate the way information passes through synaptic cleft and terminals [1-4]. As a matter of fact, various aspects and parts of bio-communication systems have been modeled and dealt with during the last decade.

Theoretical considerations about single neuron information capacity have been regarded to quantify magnitude of information a neuron carries. Therefore, upper bound of information capacity in neural response has been calculated by mathematical formulation. Another integral factor in neural signaling is the pre-synaptic spiking sequence pattern. In [5, 6], the relations between membrane potential variation of pre-synaptic and calcium ions concentration in terminal is explained. This variation can totally affect the vesicle release probability (discussed in the next section). The biological mechanism of such a process is expressed in [7]. There have been many different approaches to model neural activity under certain excitations. One simple but physically intuitive one is known as leaky integrated-and-fire (LIF) model. LIF model can mimic the behavior of the biological neuron with minimum number of circuit element unlike other models [8]. It characterizes the action potential occurrence as a charge and discharge process of a RC circuit. The time which neurons resist to respond after an action potential is called refractory period. This time duration is limits the upper excitation frequency of neurons (assumed to be 2 ms in this paper).

Fig. 1 shows an example of a neurotransmission in synaptic communication. The main biological process which matters in our study is threefold:

*1)* Incoming action potentials depolarizes the pre-synaptic neuron's membrane. The intracellular calcium concentration rises in this area and lead to neuro-transmitter release by a time-varying release probability.

*2)* Neuro-transmitters pass through the cleft by neuro-transmitter propagation probability.

*3) Neuro-transmitter binding probability to post-synaptic terminals* shows the magnitude of binding neuro-transmitters to some receptors at the output terminal.

In this paper, effects of excitation frequency change on synaptic signals are investigated. The pre-synaptic neuron's spikes are generated by a LIF model which shows a low-pass behavior of neural networks. Cut-off frequency of this filter is computed for upper bound frequency excitation. Furthermore, the post-synaptic neuron's spiking rate is calculated by changing the frequency of external stimulation. In the second section, biological concepts, formulation and computation for vesicle release probability and LIF model is presented. We commence the section three with system design block diagram and the way to derive the output spiking rate based on the biological process presented in the first two sections. After that, numerical results are displayed and analyzed in the fourth section. The last section covers the conclusion and final notions about benefits of this research.

## II. IMPORTANT CONCEPETS OF SYNAPTIC COMMUNICATION AND LIF MODEL

In this section, the mathematical formulation of synaptic mechanism of neurons is presented. All the parameters of this section's formulas and their assigned values are listed in table 1.

### A. Vesicle Release

During action-potential-triggered transmission, vesicle release happens when $Ca^{+2}$ enters pre-synaptic neuron via voltage-sensitive ion channels [9]. In this condition, at least four calcium ions must cooperate to activate the release site. These ions are modeled as four independent gates ($S_1$-$S_4$) with $O_i$ as the opening probability of each gate. The calcium concentration at the release site is found by voltage membrane.

$$Ca^{+2} = -A \cdot g_{ca} \cdot P \cdot \frac{2FV}{RT} \cdot [\frac{Ca_{ex}}{1-\exp(2FV/RT)}] \quad (1)$$

The equation 1 has an extremum at point $V=0$. In the last two sections, this critical point is explained in terms of dynamical behavior of neural activity and the corresponding LIF model. In equation 2, the differential equation of open gates is shown. Thus, open gates form stochastic processes.

$$\frac{dO_i}{dt} = k_i^+ Ca^{+2} - (O_i \cdot (k_i^+ Ca^{+2} + k_i^-)); i = 1,2,3,4 \quad (2)$$

And finally equation 3 aids to compute the vesicle release probability. This formula marks all four active channels.

$$P_{rel}(t) = \prod_{i=1}^{4} O_i(t) \quad (3)$$

### B. LIF Model

In the LIF model, neuron is treated like a parallel combination of a "leaky" resistor (conductance, $G_L$) and a capacitor (C) [8]. When membrane's potential exceeds threshold, the capacitor discharges until the membrane's voltage reaches the resting potential $V_r$ by a little swing. In this regard, equation 4 is sufficient to account for the explained system:

$$C\frac{dV}{dt} = -G_L(V(t)-V_r) + I(t) \quad (4)$$

### C. Spontaneous Release Rate

Spontaneous release rate ($\lambda_0$) at the pre-synaptic terminal can occur when the pre-synaptic membrane is not depolarized [10]. This rate is not caused by the excitation and is independent of the probability of release. As a deduction, it is usually considered as a noisy element in output information membranes. Its value is nonlinearly related to the calcium concentration:

$$\lambda_0(t) = a_3(1+\exp(\frac{a_1-Ca^{+2}(t)}{a_2}))^{-1} \quad (5)$$

Note that the equations are applicable when the mentioned steps in the last section are independent of each other.

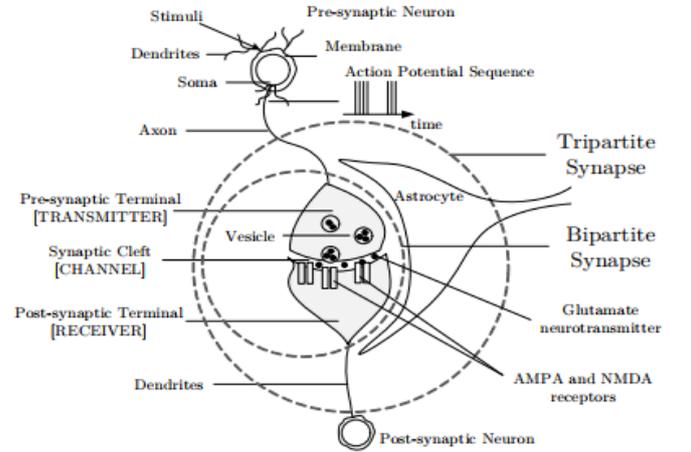

Fig. 1. Schematic diagram of the neuronal anatomy with bipartite and tripartite synapses. Image is copied from [10].

## D. Single Frequency External Excitation

The external excitation affects the model as a leaky current. According to the Kramers-Kronig equations, when there is no frequency in the spectrum but a single one, image components are zero. Therefore, there will be no imaginary part of dielectric properties. As a result, if the excitation is single frequency, then the body is like a medium with constant dielectric properties for the incident wave. Thus, the incident and penetrated wave have the same frequency ($f_e$) and the leaky current can be modeled by a simple sinusoidal signal (such as $\cos(2\pi f_e t)$).

TABLE I. PARAMETERS TERMINOLOGY AND VALUES OF THE FIRST FOUR EQUATIONS

| Parameter | Symbol | Value in this paper |
|---|---|---|
| Single Channel Conductance | $g_{ca}$ | 12 pS |
| Concentration to membrane potential converter factor | P | 1.6 mV/mM |
| Thermal Voltage | $\frac{RT}{F}$ | 26.7 mV |
| – | A | 0.1 mM/fA |
| External $Ca^{+2}$ concentration | $Ca_{ex}$ | 10 mM |
| Membrane Voltage | V | Time-Varying |
| Opening Rates | $k_1^+$ | $3.75 \times 10^{-3}$ ms$^{-1} \cdot \mu$M$^{-1}$ |
| Opening Rates | $k_2^+$ | $2.5 \times 10^{-3}$ ms$^{-1} \cdot \mu$M$^{-1}$ |
| Opening Rates | $k_3^+$ | $5 \times 10^{-4}$ ms$^{-1} \cdot \mu$M$^{-1}$ |
| Opening Rates | $k_4^+$ | $7.5 \times 10^{-3}$ ms$^{-1} \cdot \mu$M$^{-1}$ |
| Closing Rates | $k_1^-$ | $4 \times 10^{-4}$ ms$^{-1}$ |
| Closing Rates | $k_2^-$ | $10^{-3}$ ms$^{-1}$ |
| Closing Rates | $k_3^-$ | 0.1 ms$^{-1}$ |
| Closing Rates | $k_4^-$ | 10 ms$^{-1}$ |
| Resting Potential | $V_r$ | -50 mV |
| Conducatance in LIF | $G_L$ | .01 S |
| Capacitance in LIF | C | 100 F |
| Threshold potential | $V_{th}$ | 10 V |
| – | $a_1$ | 7181 μM |
| – | $a_2$ | 606 μM |
| – | $a_3$ | 100 ms$^{-1}$ |

## III. SYSTEM MODEL

The firing rate of the output is consequence of the both spontaneous vesicle release and the stimulus one. The vesicle release rate ($\lambda_1$) caused by the stimulus needs the polarization of the membrane voltage. So, the input spiking rate is multiplied by the vesicle release probability to let the neuro-transmitter comes into synaptic cleft. Among the three mentioned processes in the second section, neural communication is mainly constrained by release probability. Accordingly (like most of the papers in this area), propagation and binding probabilities are assumed to be unity in this research. Henceforth, by equation 5, we can compute the output spiking rate ($\lambda_0$):

$$\lambda_2(t) = \lambda_0(t) + P_{rel}(t)\lambda_1(t) \quad (5)$$

The systematic block diagram of the simulation in this paper is depicted in Fig. 2. It sheds light the procedure needed to generate $\lambda_2$. The whole frequency dependency of the scheme is obvious at the first stage.

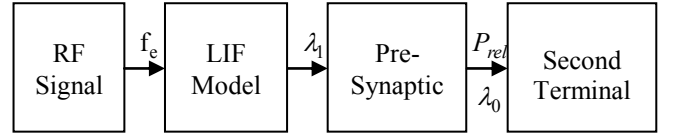

Fig. 2. Block diagram of the neuronal communication process to reach output firing rate at the post-synaptic neuron.

## IV. NUMERICAL RESULTS

This section presents the outcomes of the mentioned system model simulation with values of table 1. The effects of changing frequency excitation on the whole synaptic communication system is evaluated and analyzed.

Fig. 3 illustrates the graphs about mean value of input voltage membrane and probability of release together. Both figures are drawn in normalized scale for a better intuitive comparison. Equation 1 has a extremum when membrane voltage is zero. It is noticeable that the mean value of voltage membrane owns the same behavior as instantaneous one. The developed voltage membrane is outcome of first order LIF model. Its eigenvalue (first derivative) is zero around 200 Hz. Also, at this frequency, second derivative is non-zero. Besides, external excitation in LIF model ($\cos(2\pi f_e t)$) is not function of the state variables. For the mentioned three-fold reasons, when EM-wave is excited around 200 Hz, there happens to be a fold bifurcation in its node equilibrium point.

Fig. 4 depicts the cut-off frequency by altering the time constant of LIF circuit ($\tau = RC$). Based on this graph, the upper bound of excitatory frequency for some medical devices can be attained. So, such a figure can be very useful for engineers to design neural imaging and non-invasive excitation systems.

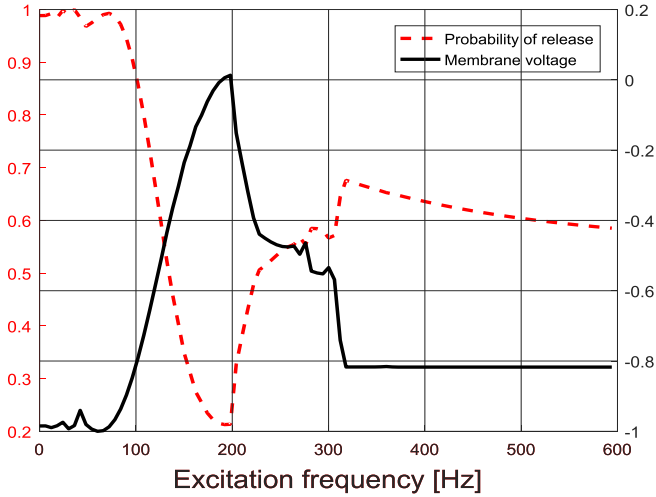

Fig. 3. Normalized values of probability release and mean value of membrane voltage are sketched by changing excitation frequency.

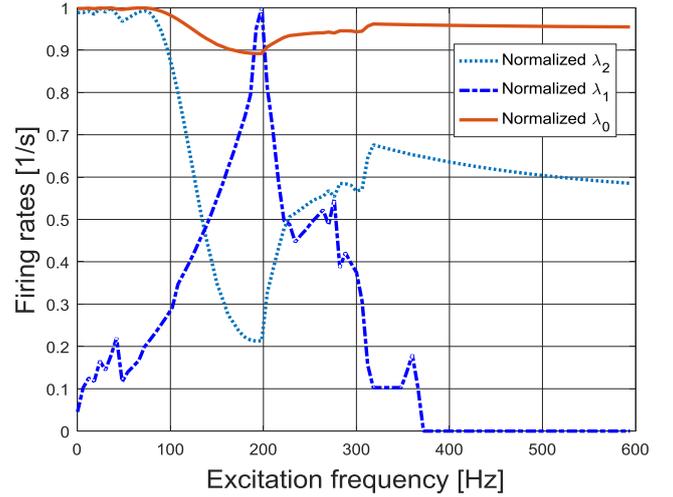

Fig. 5. Input and output firing rates based on different excitation frequencies.

According to Fig. 5, post-synaptic firing rate can be non-zero, even if the input excitatory one vanishes at the cut-off frequency. This is due to the spontaneous vesicle release. It is remarkable that even $\lambda_0$ alters with calcium influx and the excitation frequency. The peak input spiking rate is at the equilibrium point. However, the small value of release probability at this frequency causes the maximum post-synaptic firing rate occurs at another one (near 310 Hz).

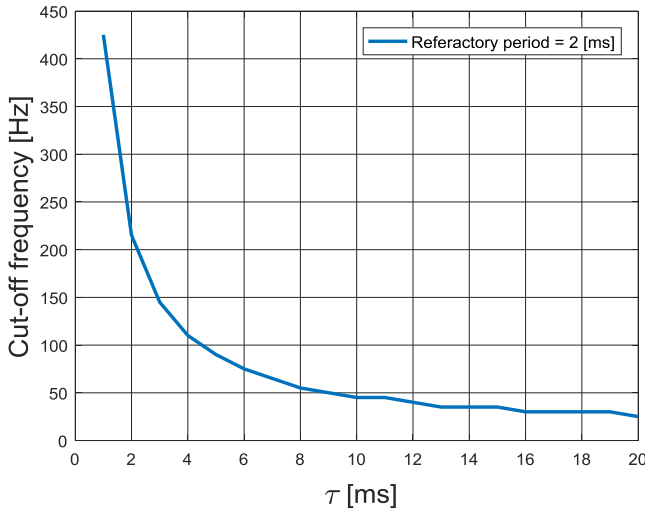

Fig. 4. Cut-off frequency for excitation of the pre-synaptic neuron versus the time constant chnages of LIF model.

## V. Conclusion

In this paper, importance of proper frequency selection for external excitation of neurons is investigated. It is shown that there is a cut-off frequency for any neural network working like a low-pass filter. This frequency is affected by biological features of neural system that is modeled as RC circuit's passive elements' values. It is intuitive that mean value of the membrane voltage can display the equilibrium point of spiking rate near the frequency that is expected from instantaneous voltage. Another plausible thing is about the factors that can change this specific point, because the dynamic behavior of the system alters at this frequency. This interesting concept opens the way for further exploration about dynamical systems relations with external wave excitations.

Although after the cut-off frequency there are no spikes in the input membrane's voltage, but there still can be firing rates in the post-synaptic neuron. This happens due to non-zero values of $\lambda_0$ at those frequencies. This value is also related to the calcium flux that can be affected by excitation either. Hence, we can have excitations that do not cause firing rates at the input, but can alter the neural communication. This is a remarkable point for further researches about sub-threshold neural activities and the way they affect the information rate of brain. These sub-threshold voltage patterns may help to heighten the frequency of external excitation to the ones bigger than cut-off frequency.

Another outcome of this paper is that highest input spiking rate does not necessarily have the same frequency with hugest information rate. In fact, the calcium flux and accordingly probability of release may cause that the peak output firing rate occurs at another frequency.